\documentclass[fp]{jpsj3}
\usepackage{txfonts}
\usepackage{url}

\title{Partial Pressures in Liquid Mixtures and Osmotic Pressures}

\author{Junzo Chihara$^1$\thanks{jrchihara@nifty.com} and Mitsuru Yamagiwa$^2$}
\inst{$^1$Higashi-ishikawa 1181-78, Hitachinaka, Ibaraki 312-0052, Japan \\
$^2$Kansai Photon Science Institute, JAEA, Kizugawa, Kyoto 619-0215, Japan}

\abst{When an osmotic system is composed of 1- and 0-species particles, which are confined to the volumes, $V\equiv V_{\rm a}+V_{\rm b}$ and $V_{\rm b}$, by the wall pressures,  $P_1^{{\rm w}_{\rm a}}$  and $P_0^{{\rm w}_{\rm b}}$, respectively, 
we prove a law of partial pressures as described in the forms:  $P_1^{\rm w_{\rm a}}=P_{\rm 1b}+\Gamma$, $P_0^{\rm w_{\rm b}}=P_{\rm 0b}-\Gamma$ and $P=P_0^{\rm w_{\rm b}} + P_1^{\rm w_{\rm a}}= P_{\rm 0b}+P_{\rm 1b}$ for the total pressure $P$. Here, the partial pressures, $P_{\rm \alpha b}$, are given by the virial equation, and the pressure difference $\Gamma$ on a semipermeable membrane appears owing to the presence of density discontinuity at the membrane. As a result, we show that the partial pressures defined by the wall pressures satisfy the law of partial pressures and are measurable in liquid mixtures confined, also, in a finite volume. Furthermore, the partial pressures are shown to play an important role in treating the gas-liquid equilibrium as well as osmotic systems: this equilibrium in mixtures is established by the balance of each partial-pressure for every component between the two phases. Hence, for a dilute solution Henry's law,  Raoult's law and van't Hoff's law can be derived on the basis of the concept of partial pressure.
}

\kword{Osmosis, partial pressure, liquid mixture, gas-liquid equilibrium}
\begin{document}
\maketitle

\section{Introduction}\label{sec:int}
At the present stage, it is a common belief that the law of partial pressures is only applicable to ideal gases as Dalton's law. We have proven that the total pressure of the electron-nucleus mixture such as liquid metals and plasmas can be represented as the sum of the electron and nuclear pressures, which are defined by the wall potentials confining the electrons and nuclei in the finite volume, respectively \cite{ChiharaEN,ChiharaUn}.  
We explore the meaning of this definition to investigate here the problem of \lq partial pressure' for liquid and real gas mixtures. 

Let us consider an osmotic system: a sugar-solution compartment separated from a pure-water compartment by a semipermeable membrane, for example. 
The osmotic pressure, that is, the pressure on the membrane exerted by the sugar is different from the internal sugar pressure in the solution \cite{Itano}. This fact may give rise to a confusion that the law of partial pressures can not be applied to a
sugar-water mixture, as is found in an erroneous definition that the water partial pressure is defined to be the wall pressure confining the water on the pure water side and the sugar pressure to be the membrane pressure exerted by the sugar in the osmotic system.\cite{HOBBIE,Woo}

On the other hand, the term, partial pressure of oxygen in the blood for example, is used commonly in physiology, where the partial pressure of a gas dissolved in a liquid is defined as the partial pressure that the gas would exert if the gas phase were in equilibrium with the liquid.\cite{Schmidt-Nielsen,Das,PP,Stewart} Many people\cite{Willmer,Hoar,Watten,Das} believe that this definition stems from the fact that the partial pressure of every component ($\alpha$) in the two phases (gaseous[G] and liquid[L]) is the same at equilibrium, that is, $P_{\alpha G}\!=\!P_{\alpha L}$. However, this statement is not true: the gas-liquid phase equilibrium in mixtures is established by the balance of each partial-pressure for every component between the two phases, but its partial pressure is not the same, $P_{\alpha G}\!=\!P_{\alpha L}\!+\!\Gamma_{\alpha}$, as will be discussed in this paper. Note that this relation is meaningless if we cannot define the partial pressures, $P_{\alpha L}$, in the liquid phase. 
Moreover, the definition in physiology produces an absurd example that the sugar pressure is zero in the sugar solution, since the sugar is non-volatile.

In this investigation, in the first place, we show that, in general, for a mixture confined by the wall to a finite volume, the partial pressure of each component can be defined by the pressure on the wall exerted by each component, and the total pressure is represented by the sum of these partial pressures. 
In the second place, on the basis of first principles (the virial theorem) we prove in the general situation 
an experimental fact found in the molecular dynamics simulation performed by Itano {\it et.\ al.} \cite{Itano};
that is, the fact that the "partial" pressure of the solute (such as sugar) 
is higher than the semi-membrane pressure confining the solute by a difference $\Gamma$, and 
the solvent (water) pressure in the solution is lower than the wall 
pressure confining the solvent by the same difference $\Gamma$ in the osmotic system.
In the third place, the partial pressures are shown to be measurable and important 
physical quantities by using the results found in osmotic systems; as a consequence, the law of partial pressures for interacting systems is proven as a significant physical law.
In addition, the gas-liquid equilibrium is shown to be a special state
 of an osmotic system, which is described in terms of partial pressures, $P_{\alpha G}\!=\!P_{\alpha L}\!+\!\Gamma_{\alpha}$; also we show that this relation leads to Henry's law and Raoult's law in the dilute limit.

When we take the sugar solution and the pure water separated by the membrane as a whole one system confined to a finite volume, there is a step-function-like jump of the sugar-density on the membrane, since the sugar exists only in the solution. 
 From this, we find that internal partial pressures of sugar and water are related to 
the osmotic and wall pressures in terms of the pressure difference caused by the density discontinuity.  This view is based on the concept of partial pressure. 
%On the other hand, without use of the partial pressures, Lion and Allen\cite{Lion} des%cribed the osmotic pressure using the virial theorem in a similar way to ours. 
The concept of partial pressure enables us to see the physical meaning of their virial expression for the osmotic pressure as shown later.

\section{Partial Pressures and Osmotic Pressures}\label{s:OSMmain}

At the beginning, we write up the basic equations and the concept "wall pressure", necessary 
to derive the pressure relations in this section.

\noindent(I) {\bf Virial theorem for one particle}:

In the system confined by the wall potential $U^{\rm w}$ to the volume $V$ at a temperature $T$, where the particles are interacting via a binary potential $v_{ij}$ under the external potential $U^{\rm ex}$, 
the virial theorem for one arbitrary $i$-particle \cite{ChiharaEN} is written as
\begin{equation}
2\langle {p}_i^2/2m_i\rangle=\langle{\bf r}_i\cdot\nabla_i(U+U^{\rm w})\rangle=3k_{\rm B}T\,. \label{e:vE}
\end{equation}
Here, ${\bf p}_i$ and ${\bf r}_i$ are the momentum and position of $i$-particle, respectively, and $U \equiv \sum_{i<j}v_{ij}+U^{\rm ex}$. In the above, the bracket $\langle\rangle$ denotes the ensemble average. Equation (\ref{e:vE}) results from the virial theorem: 
 $\lim_{t\rightarrow \infty} {1 \over t}\int_{t_0}^{t_0+t}{d({\bf p}_i\cdot{\bf r}_i) \over dt}dt=0$.
 
\noindent(II) {\bf Wall pressure}:

When a fluid  is confined by the wall pressure $P^{\rm w}$ to a finite volume $V$, the wall pressure $P^{\rm w}$  is in balance with the hydrostatic pressure $P$: 
 $P^{\rm w} = P$ just at the surface, thermodynamically; an external force $F$ on the wall with a surface $S$ provides the wall pressure $P^{\rm w}=F/S$. 
Note that this relation is derived from a fundamental assumption based on 
the virial theorem (\ref{e:vE}) as shown below. The wall pressure $P^{\rm w}$ caused by the wall potential $U^{\rm w}$ on the particles is defined by 
\begin{equation}\label{e:wallP}
-\oint_{\partial V}P^{\rm w}{\bf r}\cdot d{\bf S}\equiv -\sum_{i \in V}\langle{\bf r}_i\cdot\nabla_i{U}^{\rm w}\rangle=\sum_{i \in V}\langle{\bf r}_i\cdot{\bf F}^{\rm w}_i\rangle \,.
\end{equation}
The virial theorem (\ref{e:vE}) provides the basis for the standard assumption concerning the relation between the force exerted by the wall on the particles and the hydrostatic pressure $P$:
\begin{equation}\label{e:gVir}
\oint_{\partial V}P^{\rm w}{\bf r}\cdot d{\bf S}=\oint_{\partial V}P{\bf r}\cdot d{\bf S}
= \int_V ( 3 {P} +{\bf r}\cdot\nabla {P}) d{\bf r}\,,
\end{equation}
that is, 
\begin{equation}\label{e:gVir2}
P^{\rm w}=P=\sum_{i \in V}\left[\, 2\langle \frac{p_i^2}{2m_i}\rangle-\langle{\bf r}_i\cdot\nabla_i{U}\rangle \right]/{3V}\,,
\end{equation}
when the external potential $U^{\rm ex}$ becomes zero near the boundary surface $\partial V$ to provide a constant pressure on $\partial V$. Hereafter, we denote the surface of a volume $V$ by the symbol $\partial V$, and a pressure P denotes a function of the coordinate, when it is involved in the integrand of the surface or volume integrals.
Usually, the wall pressure $P^{\rm w}$ is {\it omitted} in the virial equation in the above, but it is important in
treating osmotic systems. When the wall with a surface $L^2$ is movable such as a membrane or a piston, an external force $F$ on the wall provides 
the wall pressure $P^{\rm w}=F/L^2$ to keep the wall at rest. 
Therefore, it should be kept in mind that the wall pressure $P^{\rm w}$ represents also 
the experimentally {\it measurable} external pressure, as well as the pressure exerted by the fluid on the wall in conjunction with the wall pressure acting on the fluid.
Thermodynamically, the wall potential $U^{\rm w}$ is assumed to be perfectly elastic and becomes abruptly infinite at the surface ${\partial V}$, and hence the density of the system becomes uniform from just inside the wall.

\noindent(III) {\bf The virial equation for an arbitrary volume $\Omega$ in the system}:

The virial equation is expressed by (\ref{e:gVir2}), and written for an arbitrary macroscopic volume $\Omega$ with a constant pressure-surface $\partial \Omega$ in $V$ in the form \cite{More79,Chihara01,BaderAus,Bader}:
\begin{eqnarray} 
\oint_{\partial \Omega}{P}{\bf r}\cdot d{\bf S}
=\sum_{i \in \Omega}\,\left[\, 2\langle \frac{p_i^2}{2m_i}\rangle-\langle{\bf r}_i\cdot\nabla_i{U}\rangle \right]\,. \label{eq:1}
\end{eqnarray}

\noindent(IV) {\bf Discontinuous pressure:}

In the case where the pressure ${P}$ has a uniform but different pressure $P_i$ in each of the two domains, ${V_a}$ and ${V_b}$ ($V\!=\!{V_a}\!+\!{V_b}$), separated by a surface $S(\!={\partial V_b})$ of the volume ${V_b}$ involved in the volume $V$, we obtain the following relation:
\begin{eqnarray}
\oint_{\partial V}\!{P}{\bf r}\cdot d{\bf S}
&=& \oint_{\partial V_a}\!{P}{\bf r}\cdot d{\bf S}
 + \oint_{\partial V_b}\!{P}{\bf r}\cdot d{\bf S}
 + [P(S^{\rm out})\! -\! P(S^{\rm in}) ]\oint_{\partial V_b}\!\!\!{\bf r}\cdot d{\bf S} \,. \label{e:avD}
\end{eqnarray}
Here, $P(S^{\rm out}) \equiv \lim_{\epsilon\rightarrow 0}P(S_{\!+\!\epsilon})$ and $P(S^{\rm in}) \equiv \lim_{\epsilon\rightarrow 0}P(S_{\!-\!\epsilon})$.
 Symbols, $S_{\!+\!\epsilon}$ and $S_{\!-\!\epsilon}$, denote a surface $S$ shifted 
outside by $\epsilon$ and that shifted inside, respectively. In this case the surface integral $\oint_{\partial V}\!{P}{\bf r}\cdot d{\bf S}$ cannot be calculated by the volume integral $\int_{V}\!\nabla [{P}{\bf r}] dV$ due to the discontinuity of pressure in the volume. Instead of using the volume integral, we can evaluate the surface integral by using the identical relation:
$\oint_{\partial V}\!{P}{\bf r}\!\cdot\! d{\bf S}
=[\oint_{\partial V}\!{P}{\bf r}\!\cdot\! d{\bf S}-\oint_{S_{\!+\!\epsilon}}\!{P}{\bf r}\!\cdot\! d{\bf S}]+[\oint_{S_{\!+\!\epsilon}}\!{P}{\bf r}\!\cdot\! d{\bf S}-\oint_{S_{\!-\!\epsilon}}\!{P}{\bf r}\!\cdot\! d{\bf S}]
+\oint_{S_{\!-\!\epsilon}}\!{P}{\bf r}\!\cdot\! d{\bf S}$, 
which leads to (\ref{e:avD}) in the limit $\epsilon\rightarrow 0$, with help of 
$\lim_{\epsilon\rightarrow 0}[\oint_{\partial V}\!{P}{\bf r}\!\cdot\! d{\bf S}-\oint_{S_{\!+\!\epsilon}}\!{P}{\bf r}\!\cdot\! d{\bf S}]=\int_{\partial V_{\rm a}} P{\bf r}\!\cdot\! d{\bf S}=3P_{\rm a}V_{\rm a}$.\\
 
Hereafter, we derive new pressure relations using the relations in the above (I)--(IV) as already established ones.
In the first place, we define partial pressures for a mixture composed of two types of species, 0 and 1, is confined  by wall potentials, ${U}_0^{\rm w}$ and ${U}_1^{\rm w}$, for each component to a cubic volume $L^3(\!=\!V)$; a corner is put at the origin of an orthogonal coordinate system.
The total pressure is caused by the impulses of particles colliding with the wall, which is assumed to be perfectly reflective. As a result, it is important to note that the contribution of {\it single} particle to the total pressure can be defined even in a liquid as the same as in an ideal gas.\cite{RKubo} Here, let us consider how single particle-{\it i} contributes the pressure on the wall located at $x=L$
with an area $L^2$. When the particle-{\it i} collides with the $L^2$-wall at time $t'$ between $t$ and $t+\Delta t$, this particle produces an impulse to the wall:
\begin{equation}
\Delta {\bf p}_i^{\rm w} \equiv {\bf p}_i(t+\Delta t)-{\bf p}_i(t)
={\bf F}_i({\bf r}_i(t'))\Delta t
=  \nabla_i U_\alpha^{\rm w}({\bf r}_i(t'))\Delta t \,.
\end{equation}
Therefore, we obtain the following equation if $\Delta t$ is sufficiently small:
\begin{eqnarray}
{\Delta {\bf p}_i^{\rm w} \over \Delta t}&=&{\bf F}_i({\bf r}_i(t))\\
&=&{\partial  U_\alpha^{\rm w}({\bf r}_i(t))\over \partial x_i}{\bf e}_x 
= \left\{
 \begin{array}{ll}
{2|p_{ix}| \over \Delta t} {\bf e}_x& \text{for a collision with $L^2$ during $\Delta t$ } \\
 0    & \text{for no collision with $L^2$ during $\Delta t$ }.
  \end{array}
 \right.
\end{eqnarray}
Since the time average of impulses per unit time (that is, the sum of impulses per unit time) provides a force on the surface $L^2$ exerted by the particle-{\it i}, the contribution of this {\it single} particle  to the pressure, $P_i$$\,$, is written as the same as in an ideal gas\cite{RKubo} in the form
\begin{equation}
P_i=\left\langle{\Delta {\bf p}_i^{\rm w} \over \Delta t}\cdot{\bf e}_x\right\rangle_{\rm time}/L^2=\langle {\bf F}_i({\bf r}_i(t))\cdot {\bf e}_x\rangle_{\rm time}/L^2=\langle {\bf F}_i({\bf r}_i)\cdot {\bf e}_x\rangle_{\rm ensemble}/L^2\,. \label{1mf}
\end{equation} 
In the above, the time average is replaced by the ensemble average. In final, the partial pressure of $\alpha$-species can be expressed by the sum of each pressure of single particle belonging to the $\alpha$-species,
\begin{eqnarray}
P_{\rm \alpha}&=&\sum_{i \in \alpha}\left\langle{\Delta {\bf p}_i^{\rm w} \over \Delta t}\cdot{\bf e}_x\right\rangle_{\rm time}/L^2=\sum_{i \in \alpha}\langle{{\partial U^{\rm w}_\alpha \over \partial x_i}\rangle }/L^2 \label{e:dPP1} \\
&=&{1 \over 3L^3}\sum_{i \in \alpha}\left[2\langle {p_i^2 \over 2m_{\rm \alpha} }  \rangle - \langle {\bf r}_i\cdot\nabla_i U \rangle \right]\,. \label{pp2}
\end{eqnarray}
In this derivation we have used the following relation:
\begin{equation}
\left\langle x_i{\partial U^{\rm w}_\alpha \over \partial x_i}\right\rangle=L\left\langle {\partial U^{\rm w}_\alpha \over \partial x_i}\right\rangle ={1 \over 3}\left[ 2\langle {p_i^2 \over 2m_\alpha}\rangle - \langle {\bf r}_i\cdot\nabla_i U \rangle \right]\,, \label{1vt}
\end{equation}
which is based on the fact that ${\partial U^{\rm w}_\alpha \over \partial x_i} \neq 0$ only for 
$x_i=L$ and the pressure is isotropic with help of the virial theorem (\ref{e:vE}).
Equation (\ref{e:dPP1}) indicates that the partial pressure $P_{\rm \alpha}$ can be defined generally in terms of the wall potential $U_\alpha^{\rm w}$  in the same way to the total pressure (\ref{e:gVir}) as follows:
\begin{equation}
3VP^{\rm w}_\alpha \equiv \sum_{i \in \alpha}\langle{\bf r}_i\cdot\nabla_i{U}_\alpha^{\rm w}\rangle = \oint_{\partial V}P_\alpha{\bf r}\cdot d{\bf S}=\sum_{i \in \alpha}\left[2\langle {p_i^2 \over 2m_{\rm \alpha} }  \rangle - \langle {\bf r}_i\cdot\nabla_i U \rangle \right]  \,. \label{e:dPPbW}
\end{equation}
 Thus, when the interatomic interaction is via binary potentials $v_{\alpha \beta}(r)$, the partial pressure $P_\alpha$ can be written in the form of the virial equation \cite{Hansen}:
\begin{eqnarray}\label{e:Pvir1}
P^{\rm w}_\alpha=P_\alpha& = &k_{\rm B}T \rho_\alpha
-\frac16 \rho_\alpha^2\int r\frac{dv_{\alpha\alpha}(r)}{dr}g_{\alpha\alpha}(r)d{\bf r}
+\sum_{i\in \alpha}\langle{\bf r}_i\cdot{\bf F}_i^\alpha \rangle/3V \,, 
\end{eqnarray}
in terms of the radial distribution functions $g_{\alpha \beta}(r)$ and the density $\rho_\alpha$ of $\alpha$-component with pair interactions $v_{\alpha \beta}(r)$.
Here, ${\bf F}_i^\alpha $ indicates the total force on i-particle of $\alpha$-species exerted by all particles of different species ($\bar \alpha$) to $\alpha$, defined by 
${\bf F}_i^\alpha\equiv -\sum_{l\in \bar\alpha}\nabla_i v_{01}(|{\bf r}_i-{\bf r}_l|)$, 
and these forces satisfy the following relation:
\begin{equation}\label{e:sumF01}
\sum_{i\in 0}\langle{\bf r}_i\cdot{\bf F}_i^0 \rangle
+\sum_{l\in 1}\langle{\bf r}_l\cdot{\bf F}_l^1 \rangle
=-V\rho_0\rho_1\int r\frac{dv_{01}(r)}{dr}g_{01}(r)d{\bf r}\,.
\end{equation} 
Also, the total pressure $P$ is the sum of partial pressures $P_\alpha$  defined by (\ref{e:dPP1}): $P=P_{\rm 0}+P_{\rm 1}$.  This is referred to as the law of partial pressures.

%\begin{wrapfigure}{l}{6.6cm} % l: LEFT, 6.6cm: WIDTH
\begin{figure}
\centerline{\includegraphics[width=5.5 cm,height=3.5 cm]{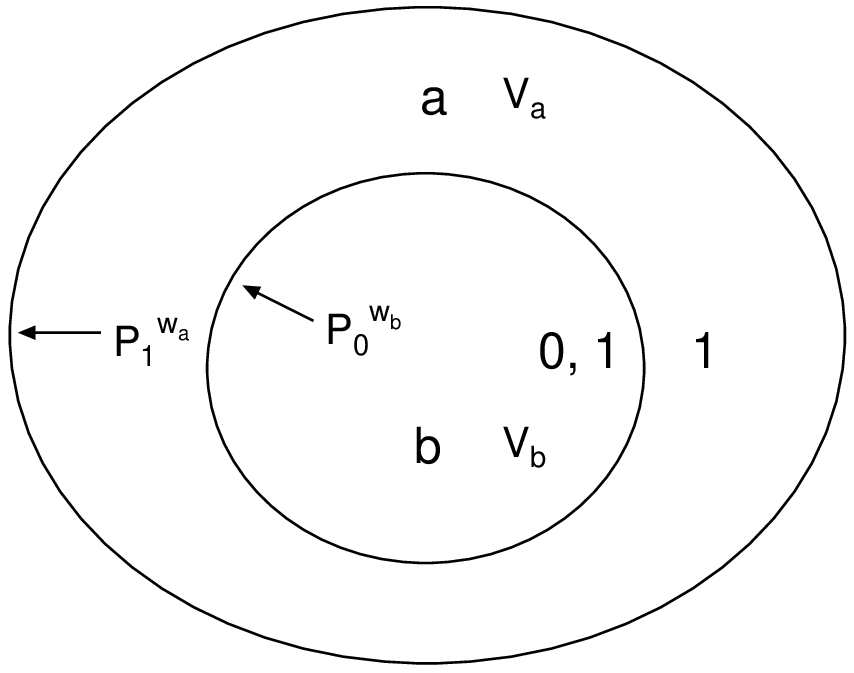}}
\caption{An osmotic system composed of 1- and 0-species particles, which are confined to the volumes, $V\equiv V_{\rm a}+V_{\rm b}$ and $V_{\rm b}$, by the wall pressures,  $P_1^{{\rm w}_{\rm a}}$ and $P_0^{{\rm w}_{\rm b}}$, respectively.}
\label{fig-1}
\end{figure}
%\end{wrapfigure}

Next, we examine the relation between the osmotic pressure and the partial pressure in a solution. Let us consider a system where the 1-species particles and the 0-species particles are confined to the volume $V$ and the volume $V_{\rm b}$ (contained in $V$) by the wall pressures, ${P}_1^{\rm w_{\rm a}}$ on $\partial V$ and ${P}_0^{\rm w_{\rm b}}$ on $\partial V_{\rm b}$, respectively, as shown in Fig.~\ref{fig-1} with $V_{\rm a}\equiv V-V_{\rm b}$. Here, the wall $\rm w_{\rm b}$ becomes equivalent to a semipermeable membrane. In thermodynamics, the volumes, $V$ and $V_{\rm b}$, are defined by using the wall potentials which are perfectly elastic and become abruptly infinite at surfaces. Therefore in this system, the density and pressure of 0-species particles become zero in $V_{\rm a}$ and jump to $\rho_{0b}$ and $P_{0b}$ in $V_{\rm b}$, respectively, with the boundary $S_{\rm b}(\equiv {\partial {\rm V}_b)}$ forming a discontinuity surface in $V$. Because of this discontinuity surface $S_{\rm b}$, the formula to calculate pressures (\ref{e:gVir}) with use of the volume integral leads to infinite divergence due to the term 
$\nabla P(\bf r)$ involved in the integrand of (\ref{e:gVir}).
The virial theorem, (\ref{eq:1}) and (\ref{e:vE}), avoiding this divergence with the help of (\ref{e:avD}) generates the relations between discontinuous quantities. In treating the osmotic system, it should be kept in mind that Eq.~(\ref{e:vE}) ensures to define 
the partial pressure of $\alpha$-component in a mixture by the sum only of {\it i} belonging to $\alpha$-component (${\it i}\in \alpha$). As this result, we can obtain the relations concerning ${P}_1^{\rm w_{\rm a}}$ and ${P}_0^{\rm w_{\rm b}}$ for each component with the help of (\ref{e:avD}) and the virial theorem below.

For the case of 1-species, the partial pressure $P_1({\bf r})$ has different uniform pressures, $P_{\rm 1b}$ and $P_{\rm 1a}$, in the inner and outer sides of the surface $S_{\rm b}$, respectively, 
due to the discontinuity of the density of 0-species on the surface $S_{\rm b}$. Therefore, equation (\ref{e:avD}) provides the following relation:
\begin{eqnarray}
\oint_{\partial {\rm V}}\!\!\!{P}_{1}^{\rm w_{\rm a}}{\bf r}\cdot d{\bf S}
&=&\oint_{\partial {\rm V}}\!\!\!{P}_1{\bf r}\cdot d{\bf S}=3{P}_1^{\rm w_{\rm a}}(V_{\rm a}+V_{\rm b})\\
&=& 3P_{1a}V_a + 3P_{1b}V_b
  + 3[P_1(S^{\rm out}_{\rm b})\! -\! P_1(S^{\rm in}_{\rm b}) ]V_b \label{e:osmWa}\,.
\end{eqnarray}
%&=& \oint_{\partial {\rm V}_a}\!\!\!\!{P_1}{\bf r}\cdot d{\bf S}
% + \oint_{\partial {\rm V}_b}\!\!\!\!{P_1}{\bf r}\cdot d{\bf S}
% + [P_1(S^{\rm out}_{\rm b})\! -\! P_1(S^{\rm in}_{\rm b}) ]\oint_{\partial {\rm V}_b}%\!\!\!{\bf r}\cdot d{\bf S} \label{e:osmWa}\,.
Here, $P_1(S^{\rm in}_{\rm b})=P_{\rm 1b}$ and $P_1(S^{\rm out}_{\rm b})=P_{\rm 1a}$  are the pressures just inside and outside of the surface $S_{\rm b}$ of the volume $V_{\rm b}$, respectively. Since ${P}_1^{\rm w_{\rm a}}=P_{\rm 1a}$, the partial pressure $P_1({\bf r})$ in the a- and b-domains is related each other through the following equations:
\begin{eqnarray}
{P}_1^{\rm w_{\rm a}}&=&P_{\rm 1a}=P_{\rm 1b}+\Gamma_{\rm 1}  \label{e:Wa}\\
\Gamma_1&\equiv& P_{\rm 1a}\! -\! P_{\rm 1b} \label{e:WaG}\,,
\end{eqnarray}
which are obtained from (\ref{e:osmWa}).
The partial pressure $P_{\rm 1}(\bf r)$ becomes uniform in $V_{\rm a}$ and $V_{\rm b}$, but has a pressure difference $\Gamma_1$ between them. Here, the pressure $P_{\rm 1a}$ is determined by the virial equation for pure species-1 in the domain-a (where its pressure is given by $P_{\rm a}\!=\!P_{\rm 1a}$)
\begin{eqnarray}
P_{\rm 1a} = k_{\rm B}T \rho_{\rm 1a}
-\frac16 \rho_{\rm 1a}^2\int_{V_{\rm a}} r\frac{dv_{\rm 11}(r)}{dr}g_{1}(r)d{\bf r} \,, \label{e:Leos1}
\end{eqnarray}
which is described by the radial distribution function $g_{1}(r)$ and binary interatomic potential $v_{\rm 11}(r)$ for the domain-a, and partial pressures $P_{\alpha b}$ are given by (\ref{e:Pvir1}) for the domain-b as 
\begin{eqnarray}\label{e:Pvir-b}
P_{\rm \alpha b}& = &k_{\rm B}T \rho_{\rm \alpha b}
-\frac16 \rho_{\rm \alpha b}^2\int r\frac{dv_{\alpha\alpha}(r)}{dr}g_{\alpha\alpha}(r)d{\bf r}
+\sum_{i\in {\rm \alpha b}}\langle{\bf r}_i\cdot{\bf F}_i^{\rm \alpha b}\rangle/3V_{\rm b} \,, 
\end{eqnarray}
since there is a uniform mixture of 0- and 1-species in the domain-b, where its pressure is determined by 
$P_{\rm b}\!=\!P_{\rm 0b}+P_{\rm 1b}$.

\begin{figure}
\centerline{\includegraphics[width=5.5 cm,height=4.5 cm]{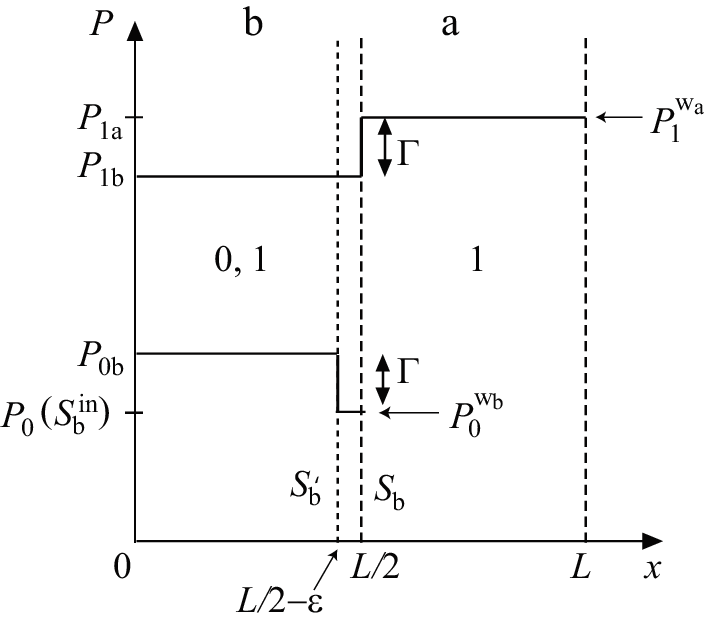}}
\caption{The pressure relations for the osmotic system for a cubic volume $L^3$ with a semipermeable membrane $S_{\rm b}$ at $x\!=\!L/2$. The b-domain involves particles of 0- and 1-species, while the a-domain contains 1-species particles only. The surface of pressure discontinuity of 0-species is shifted towards the origin by $\epsilon$ as shown by $S_{\rm b'}$.}
\label{fig-2}
\end{figure}

On the other hand for the case of 0-species, due to the density discontinuity on the surface $S_{\rm b}$(=$\partial V_{\rm b}$) in the volume $V$, the surface of pressure discontinuity mentioned above appears to be coincident with  $S_{\rm b}$. Therefore, we shift this discontinuity surface to $S_{\rm b'}$ inside of $S_{\rm b}$ by $\epsilon$ (see Fig.\ref{fig-2}), and get a final relation by taking the limit $\epsilon\rightarrow 0$ with use of $P_0(S^{\rm in}_{\rm {b'}})=P_{\rm 0b}$, $\lim_{\epsilon\rightarrow 0} P_0(S^{\rm out}_{\rm {b'}})=P_0(S^{\rm in}_{\rm {b}})={P}_0^{\rm w_{\rm b}}$ and $3V_{\rm {b'}}=\oint_{S_{\rm b'}}\!\!\!{\bf r}\cdot d{\bf S}$:
\begin{eqnarray}
\oint_{\partial {\rm V_{\rm b}}}\!\!\!\!\!\!\!{P}_0^{\rm w_{\rm b}}{\bf r}\cdot d{\bf S}
&=& 3\lim_{\epsilon\rightarrow 0}\{(S_{\rm b}\epsilon){P}_0^{\rm w_{\rm b}}\!+\!
(V_{\rm b}\!-\!S_{\rm b}\epsilon)P_{\rm 0b}\!+\!V_{\rm {b'}}[P_0(S^{\rm out}_{\rm {b'}})\! -\! P_0(S^{\rm in}_{\rm {b'}})] \}\\
&=&3V_{\rm b}P_{\rm 0b}+3V_{\rm {b}}[P_0(S^{\rm in}_{\rm {b}})\! -\! P_{\rm 0b}] \,,
\end{eqnarray}
which is derived from (\ref{e:avD}).
Here, the pressure contribution of a volume $S_{\rm b}\epsilon$ disappears in the limit $\epsilon\!\rightarrow\! 0$ and $P_0(S^{\rm in}_{\rm {b}})$ is the pressure on the wall b exerted by the 0-species particles, which is now not equal to the partial pressure of 0-species.
Hence, we get the final result:
\begin{eqnarray}
{P}_0^{\rm w_{\rm b}}&=&P_0(S^{\rm in}_{\rm {b}})=P_{\rm 0b}+\Gamma_0 \label{e:Wb}\\
\Gamma_0 &\equiv& P_0(S^{\rm in}_{\rm {b}})-P_{\rm 0b}\,.
\end{eqnarray}
The pressure $P_{\rm 0}(\bf r)$ provides ${P}_0^{\rm w_{\rm b}}$ to be equal with the osmotic pressure on the membrane $S_{\rm b}$, and has a pressure difference $\Gamma_0$ in comparison with the internal pressure $P_{\rm 0b}$ in $V_{\rm b}$.

Because the osmotic pressure $P^{\rm osm}$ is defined by the pressure difference produced across the membrane, it satisfies the following relation (see Appendix~\ref{s:osmP} for this proof):
\begin{equation}
P_0^{\rm w_{\rm b}}=P^{\rm osm}\equiv \bigl( P_{\rm 0b}+P_{\rm 1b} \bigr) - P_{\rm 1a} = P_{\rm b}-P_{\rm a} \label{e:osm1}\,,
\end{equation}
which leads to the following condition:
\begin{eqnarray}
\Gamma_0+\Gamma_1=0 \,. \label{e:cond1}
\end{eqnarray}
Owing to the relation $P_{\rm 1a}=P_1^{\rm w_{\rm a}}$, Eq.~(\ref{e:osm1}) is rewritten in the form:
\begin{equation}
P_0^{\rm w_{\rm b}} + P_1^{\rm w_{\rm a}} = P_{\rm b} = P_{\rm 0b}+P_{\rm 1b}\,.
 \label{e:mom}
\end{equation}
Thus, we have proven the relations, (\ref{e:Wa}) and (\ref{e:Wb}) with (\ref{e:cond1}), which are found numerically by the computer experiment of Itano {\it et.\ al.} with use of the perfectly reflective wall and the partial pressure definition (\ref{e:dPP1}) based on the impulses which the particles impart on the wall.\cite{Itano}.

%\begin{wrapfigure}{l}{6.6cm} % l: LEFT, 6.6cm: WIDTH
%\end{wrapfigure}

In Fig.\ref{fig-1}, the osmotic system is represented as a general volume shape; the volume shape can be deformed to the cubic volume $V=L^3$ put on the rectangular coordinates $x=y=z=L$ with a plane semipermeable membrane at $x=L/2$. The region ($0<x<L/2$) is the b-domain 
involving particles of 0- and 1-species, while the region ($L/2<x<L$) becomes the a-domain containing 1-species particles only. Naturally, the relations for ${P}_1^{\rm w_{\rm a}}$ and $P_0^{\rm w_{\rm b}}$ given by (\ref{e:Wa}) and (\ref{e:Wb}), respectively, are established for this cubic volume along the $x$-axis, as can be shown in the similar manner above. The following pressure relations are shown in Fig.\ref{fig-2} for this osmotic system: 
\begin{subequations}
\label{eq:3}
\begin{eqnarray}
P_1^{\rm w_{\rm a}}&=&P_{\rm 1b} + \Gamma   \label{e:1Gam}\\
P_0^{\rm w_{\rm b}}&=&P_{\rm 0b} - \Gamma   \label{e:0Gam}
\end{eqnarray}
\end{subequations}
with $\Gamma\equiv \Gamma_1=-\!\Gamma_0$. 
Here, note that the wall pressure $P_0^{\rm w_{\rm b}}$ for 0-species particles has a difference $\Gamma$ compared to the internal pressure $P_{0b}$ in the domain-b. 

We can understand why the pressure difference appears at the semipermeable membrane:
In a macroscopic point of view, the solutes involved in $V_b$ are pushed by the external wall pressure $P_0^{{\rm w}_{\rm 0}}=P_{\rm 0b}$ from the left side in the positive x-axis direction to support the internal pressure $P_{\rm 0b}$, and also by the external pressure $P_0^{{\rm w}_{\rm b}}$ on the semipermeable membrane from the right side in the negative x-axis direction. The solutes remain steady, if we take into account of the force on $\it i$-solute caused by the total solvents, ${\bf F}_{i\leftarrow 1}$, as follows: 
\begin{equation}\label{e:MacB}
P_0^{{\rm w}_{\rm 0}}=P_0^{{\rm w}_{\rm b}}+\sum_{i_0\in {[0,L/2]}}\langle (-{\bf e}_{\rm x})\cdot{\bf F}_{i\leftarrow 1}\rangle/L^2 = P_{\rm 0b}\,,
\end{equation}
together with the relation
\begin{equation}\label{e:GMMfij}
L^2\Gamma = \sum_{i_0\in {[0,L/2]}}\langle (-{\bf e}_{\rm x})\cdot{\bf F}_{i\leftarrow 1}\rangle
=\sum_{j_1\in {[0,L]}}\langle {\bf e}_{\rm x}\cdot{\bf F}_{j\leftarrow 0}\rangle = L^2\Delta U\,.
\end{equation}
Here, ${\bf F}_{j\leftarrow 0}$ is the force on solvent-{\it j} caused by the total solutes (0), defined by
\begin{equation}
 {\bf F}_{j\leftarrow 0}\equiv \sum_{i_0\in {[0,L/2]}}{\bf f}_{ji} 
= -\sum_{i_0\in {[0,L/2]}}{\bf f}_{ij}\,,
\end{equation}
where ${\bf f}_{ij}$ is the force exerted on particle $i$ by particle $j$, and the summation $\sum_{i_0}$ means $\sum_{i \in 0}$.
The pressure difference, $\Gamma$, defined by (\ref{e:GMMfij}) is the same expression as $\Delta U$ of Itano {\it et.\ al.} 
The density of solute molecules (0-species particles) is zero in $V_{\rm a}$ and jumps to $\rho_{0b}$ in $V_{\rm b}$, with the membrane forming a discontinuity surface in $V$. Therefore, 
the total force on the solvent, $\sum_{j_1\in {[0,L]}}\langle {\bf e}_{\rm x}\cdot{\bf F}_{j\leftarrow 0}\rangle$ , caused by all solute particles in $V_{\rm b}$ behaves like a wall potential to the solvent particles in $V_{\rm a}$, producing the pressure $\Gamma$ on the solvent particles in $V_{\rm a}$; this is the reason that the solvent pressure is higher in $V_{\rm a}$ than in $V_{\rm b}$. As the reaction to this force acting on the solvents, the solvent force $\sum_{i_0\in {[0,L/2]}}\langle (-{\bf e}_{\rm x})\cdot{\bf F}_{i\leftarrow 1}\rangle$ on the solutes in $V_{\rm b}$ supports the solute pressure $P_{\rm 0b}$ in $V_{\rm b}$ together with the confining pressure $P_0^{\rm w_{\rm b}}$ at the membrane. This is the reason that the osmotic pressure $P_0^{\rm w_{\rm b}}$ to confine the solute particles in $V_{\rm b}$ is lower than the internal solute pressure $P_{\rm 0b}$ by $\Gamma$.

Now, let us consider an osmotic system where the solute concentration $x_{\rm 0b}$ is sufficiently low; the fractional concentration is defined by $x_{\rm \alpha b}=\rho_{\rm \alpha b}/\rho_{\rm b}$. When $x_{\rm 0b}=0$, this system becomes pure solvent confined in the volume $V$, and $P_1^{\rm w_{\rm a}}$ represents the pure solvent pressure with $P_0^{\rm w_{\rm b}}=0$ owing to no solute in $V_b$. For the purpose to treat a dilute solution, we define a new quantity
\begin{equation}\label{e:dilute}
 3V_{\rm b}\Delta \equiv \sum_{i\in {\rm 0b}}\langle{\bf r}_i\cdot{\bf F}_i^{\rm 0b} \rangle-3V_{\rm b}\Gamma  \,.
\end{equation}
In terms of $\Delta$, equations, (\ref{e:1Gam}) and (\ref{e:0Gam}), are written  as follows:
\begin{eqnarray}
P_1^{\rm w_{\rm a}}&=& k_{\rm B}T \rho_{1b}
-\frac16 \rho_{1b}^2\int r\frac{dv_{11}(r)}{dr}g_{11}(r)d{\bf r}-\frac13 \rho_{0b}\rho_{1b}\int r\frac{dv_{01}(r)}{dr}g_{01}(r)d{\bf r}-\Delta\,, \label{e:1bG}\\
P_0^{\rm w_{\rm b}}&=&P_{\rm 0b}-\Gamma= k_{\rm B}T \rho_{\rm 0b}
-\frac16 \rho_{\rm 0b}^2\int r\frac{dv_{00}(r)}{dr}g_{00}(r)d{\bf r}+\Delta\,, \label{e:0bG}
\end{eqnarray}
with use of (\ref{e:Pvir-b}) and (\ref{e:sumF01}). Note that Eq.~(\ref{e:0bG}) leads to $\Delta=0$ when $x_{\rm 0b}=0$.

Here, we find that in the dilute limit Eq.~(\ref{e:0bG}) provides van't Hoff's law as a limiting law $P_0^{\rm w_{\rm b}}/x_{0b}\!=\!\lim_{x_{0b}\rightarrow 0}[P_0^{\rm w_{\rm b}}/x_{0b}]$ under the condition $\lim_{x_{0b}\rightarrow 0}[\Delta/x_{0b}]\!=\!0$.
In the dilute solution over a wide range of concentrations, computer simulation experiments \cite{Murad93,Itano} 
have demonstrated the validity of van't Hoff's law: 
\begin{equation}\label{e:OSMg}
P_0^{\rm w_{\rm b}} = P^{\rm osm} = k_{\rm B}T \rho_{0b}\,,
\end{equation}
which means that in the dilute limit the conditions, $\Delta\!=\!0$ and $\Delta/x_{0b}\!=\!0$, in (\ref{e:0bG}) are satisfied.

Furthermore, we show below that the partial pressure $P_\alpha$ even in liquids can be measurable. Note that the osmotic system is separated into two coupled systems consisting of a pure 1-fluid and a fluid mixture (1,0) when we cover the semipermeable membrane with a membrane impermeable to 1-species from the domain-a side in the osmotic equilibrium state. Also, it should be noted that the external force condition (\ref{e:PBL}) to keep the osmotic system steady holds too in the two coupled systems as shown below. These new two systems are coupled with each other in the way that the 1-species internal pressure of one system works as an external pressure to the other to sustain equilibrium. Now, the 1-species particles do not move through the semipermeable membrane and the osmotic system is changed to the separated two systems, a fluid mixture with a pressure $P_{0b}+P_{1b}$ confined in the volume $V_{\rm b}$ and a pure fluid with a pressure $P_{1a}$ in the volume $V_{\rm a}$. Therefore, the 0-species pressure on the semipermeable membrane increases by $\Gamma$ compared to the osmotic system, since $P_0^{\rm w_{\rm b}}=P_{0b}$ for this mixture in $V_{\rm b}$.
On the other hand, the membrane impermeable to 1-species covering the semipermeable membrane from the pure fluid side divides the 1-species particles by confining them in $V_{\rm a}$ and $V_{\rm b}$, and feels the pressure $\Gamma$ in the negative x direction due to the pressure difference of 1-species as shown in Fig.\ref{fig-2}.  Therefore, this membrane is pushed onto the semipermeable membrane by the pressure $\Gamma$, which is balanced to the increased pressure $\Gamma$ of the semipermeable membrane in the positive x-direction.
%As a result, this fact provides the means to measure the pressure difference $\Gamma$ involved in (\ref{e:1Gam}) and (\ref{e:0Gam}). 
Thus, in the two coupled system the partial pressures $P_{\alpha b}$ in $V_{\rm b}$ can be obtained by using values of the pressures, $P_0^{\rm w_{\rm b}}=P_{\rm 0b}$ and $P_{\rm T}^{\rm w_{\rm 0}}-P_0^{\rm w_{\rm b}}=P_{\rm 1b}$, which are observable. 
Here, $P_{\rm T}^{\rm w_{\rm 0}}$ denotes the external total wall pressure on the wall $x\!=\!0$ in Fig.~2 to support the total pressure $P_{0b}+P_{1b}$ in $V_b$
%In addition, the pressure difference $\Gamma=\Delta U$ can be calculated by evaluating the pressure difference between both sides of the membrane impermeable to 1-species with aids of the MD simulation, 
%instead of using the solute-solvent interaction force as was done by Itano {\it et. al.} \cite{Itano}.

In this way, it is shown that partial pressures, $P_{\rm \alpha b}$, are measurable using the osmotic system. As a result, the law of partial pressures (\ref{e:dPP1}) is shown meaningful even in real gases and liquids as well as in ideal gases. In other words, owing to the pressure difference $\Gamma$ on the membrane, 
the law of partial pressures for this osmotic system may be interpreted as expressed in the following forms:  
$P_1^{\rm w_{\rm a}}=P_{\rm 1b}+\Gamma$, $P_0^{\rm w_{\rm b}}=P_{\rm 0b}-\Gamma$ and $P=P_1^{\rm w_{\rm a}} + P_0^{\rm w_{\rm b}}= P_{\rm 1b}+P_{\rm 0b}$, and Eq.~(\ref{pp2}) is recovered when $V_{\rm a}=0$.

\section{Osmotic systems consisting of many kinds of species}
In \S\ref{s:OSMmain}, we treated an osmotic system consisting of two types of species, 0 and 1. Here, we increase types of permeable species such as 1, 2, ..., {\it M} in addition to impermeable 0-species in the domain-b confined by the membrane; furthermore, we  add 0-species particles in the domain-a outside the membrane. In this case, we obtain the following relations for all species except 0 $(i=1,..M)$:
\begin{eqnarray}\label{e:OSMmks1}
P_i^{\rm w_{\rm a}}&=&P_{i\rm a}=P_{i\rm b}+\Gamma_{i} \,.
\end{eqnarray}
Here, $\Gamma_i \equiv P_i(S^{\rm out}_{\rm b})\! -\! P_i(S^{\rm in}_{\rm b}) = P_{i\rm a}-P_{i\rm b}$. 
On the other hand, the following equation is established for the wall pressure ${P}_0^{\rm w_{\rm b}}$ confining the impermeable 0-species particles:
\begin{eqnarray}\label{e:OSMmks0}
{P}_0^{\rm w_{\rm b}}&=& P_{\rm 0b}- P_{\rm 0a} + \Gamma_0=P_0(S^{\rm in}_{\rm {b}})-P_{\rm 0a} \,. \label{e:Wb2}
\end{eqnarray}
Eq.~(\ref{e:Wb2}) involves $P_{\rm 0a}$ in comparison with (\ref{e:Wb}) due to the presence of the 0-species particles in the domain-a of this system. Since the difference between the pressure $P_{\rm b}$ of domain-b and the total pressure $P_{\rm a}$ in the domain-a  brings about the osmotic pressure $P^{\rm osm}= P_0^{\rm w_{\rm b}} = P_{\rm b} - P_{\rm a}$, we obtain the condition $\sum_{i=0}^{M}\Gamma_{i}=0$ for $\Gamma_{i}$, because of $P_{\rm a}\!=\!P_{\rm b}\!-\!P_{\rm 0b}\!+\!\sum_{i=1}^{M}\Gamma_{i}\!+\!P_{\rm 0a}$. On the other hand, Eq.~(\ref{e:mom}) is written in the another form: $P_0^{\rm w_{\rm b}} + \sum_{i=0}^{M}P_i^{\rm w_{\rm a}} \!=\! P_{\rm b}$.

In this situation, we consider a special state of the above osmotic system, (\ref{e:OSMmks1}) and (\ref{e:Wb2}), where the particles in the domain-a constitute the gas phase, and the particles in the domain-b keep the liquid phase \cite{Murad,Pollock} assuming that all types of species are volatile. 
In the first place, in order to maintain the solvent-0 being a liquid state, we confine the solvent-0 in the domain-b by a membrane which is {\it im}permeable to solvent-0 and permeable to the other species {\it 1,2...M}; this system is an osmotic system where the external osmotic pressure $P_0^{\rm w_{\rm b}}$ is necessary to be stable. In the second place, we add the solvent-0 in the domain-a. Here, the external osmotic pressure $P_0^{w_a}$ decreases as the density of the solvent-0 increases; at final we arrive at the state $P_0^{\rm w_{\rm b}}=0$. Because of $P_0^{\rm w_{\rm b}}=0$, we need no membrane to sustain this system. 
Since this osmotic system maintains this two-phase equilibrium even when the confining pressure $P_0^{\rm w_{\rm b}}$ of 0-species becomes zero (that is, without this membrane), we can take this state to be in the gas-liquid equilibrium; the condition, $P_0^{\rm w_{\rm b}}=0$, leads to the condition for the two-phase equilibrium $P_{\rm b}=P_{\rm a}$. 
From the condition: $P_0^{\rm w_{\rm b}}\!\!=\!P_0(S^{\rm in}_{\rm {b}})\!-\!P_{\rm 0a}\!=\!0$, there results the relation $P_0^{\rm w_{\rm a}}\!=\!P_{\rm 0a}\!=\!P_0(S^{\rm in}_{\rm {b}})\!=\!P_{0\rm b}+\Gamma_{0}$, since we need the external wall pressure $P_0^{\rm w_{\rm a}}$ on the domain-a wall to support the pressure $P_{\rm 0a}$ in the volume $V_{\rm a}$. Hence, for all species including 0-species ($i=0,...,M$), we obtain the following relation between partial pressures of the liquid- and gas-phases: 
\begin{equation}\label{e:plinGL}
P_i^{\rm w_{\rm a}}\!=\!P_{i\rm b}+\Gamma_{i}\!=\!P_{i\rm a},
\end{equation}
with the condition $\sum_{i=0}^{M}\Gamma_{i}=0$. 
Moreover, the non-volatility of $\alpha$-species can be defined by the relation: 
$P_\alpha^{\rm w_{\rm a}}\!=\! P_{\alpha\rm b}\!+\!\Gamma_{\alpha}\!=\!0$ from (\ref{e:plinGL}). 

Next, let us consider the gas-liquid equilibrium in a solution consisting of 0-solvent (liquid) and 1-solutes, where the gas-liquid equilibrium is described by (\ref{e:plinGL}) for $i=1$ and $0$:
\begin{subequations}
\label{eq:4}
\begin{eqnarray}
P_1^{\rm w_{\rm a}}&=&P_{\rm 1a}=P_{\rm 1b} + \Gamma   \label{e:1Gam2}\\
P_0^{\rm w_{\rm a}}&=&P_{\rm 0a}=P_{\rm 0b} - \Gamma\,.   \label{e:0Gam2}
\end{eqnarray}
\end{subequations}
Note that the symbol 0 denotes solvent, while the symbol 0 denotes impermeable solute in the osmotic relation, (\ref{e:1Gam}) and (\ref{e:0Gam}), since a liquid state of solvent here was sustained by the {\it im}permeable membrane to solvent at the first step. As a result the meaning of symbols 0 and 1 is exchanged here.
In the gas-liquid equilibrium, there are two approximate theories in a dilute solution; Henry's law and Raoult's law, which are valid only in the dilute limit. In contrast with these laws based on the chemical potential, Eq.(\ref{e:plinGL}) is valid without any condition since it is based on the rigorous virial theorem. 
In this problem, when the solvent becomes pure in $P_0^{\rm w_{\rm a}}$, that is, $x_{\rm 1b}\!=\!0$, there results
$P_1^{\rm w_{\rm a}}\!=\!0$. By taking account of this fact to treat a dilute solution, equations, (\ref{e:1Gam2}) and (\ref{e:0Gam2}), can be rewritten as the same as equations, (\ref{e:1Gam}) and (\ref{e:0Gam}) in the following:
\begin{eqnarray}
P_{\rm 1a}&=& k_{\rm B}T \rho_{1b}
-\frac16 \rho_{1b}^2\int r\frac{dv_{11}(r)}{dr}g_{11}(r)d{\bf r}-\frac13 \rho_{0b}\rho_{1b}\int r\frac{dv_{01}(r)}{dr}g_{01}(r)d{\bf r}-\Delta\,, \label{e:1bG2ph}\\
P_{\rm 0a}&=& k_{\rm B}T \rho_{\rm 0b}
-\frac16 \rho_{\rm 0b}^2\int r\frac{dv_{00}(r)}{dr}g_{00}(r)d{\bf r}+\Delta\,. \label{e:0bG2ph}
\end{eqnarray}
If $x_{\rm 1b}\!=\!0$, the above equations show a pure solvent-liquid state. On the basis of the above equations, let us consider 
the solubility of a gas (1-species) in a liquid (0-species) for small $x_{\rm 1b}$.

Murad and Gupta\cite{Murad} obtained Henry's constant $\kappa_{\rm H}$ by treating the gas-liquid equilibrium as the special case of an osmotic system for the solubility problem. Following their procedure [their equation (3)] we determine $\kappa_{\rm H}$ from Eq.~(\ref{e:1bG2ph}) in the form:
\begin{eqnarray}
\kappa_{\rm H}&=&\lim_{x_{1b}\rightarrow 0} \{P_{\rm 1a}/x_{1b}\}=dP_{\rm 1a}/dx_{1b}|_{x_{\rm 1b}=0}\approx  P_{\rm 1a}/x_{\rm 1b}  \label{e:Henry0} \\
&=&k_{\rm B}T \rho_{b}
-\frac13 \rho_{b}^2\int r\frac{dv_{01}(r)}{dr}g_{01}(r)d{\bf r}-\left.\frac{d\Delta}{dx_{1b}}\right|_{x_{1b}=0}\,. \label{e:Henry}
\end{eqnarray}
In the above, Eq~(\ref{e:Henry}) brings about the term , $\lim_{x_{1b} \rightarrow 0}\Delta/x_{1b}$, which is assumed to be zero as the same as for van't Hoff's law (\ref{e:OSMg}) where the solute is denoted by a symbol 0.

In the gas-liquid equilibrium, there is another law for a dilute solution: Raoult's law.
Raoult's law is a limiting law, $P_{\rm 0a}/x_{0b}\!\approx \!\lim_{x_{0b}\rightarrow 1}\{P_{0a}/x_{0b}\}$, valid only for $x_{0b}\!\approx\! 1$, 
as is Henry's law (\ref{e:Henry0}): $P_{1a}/x_{1b} \!\approx \! \lim_{x_{1b}\rightarrow 0} \{P_{1a}/x_{1b}\}$ for $x_{1b}\!\approx\! 0$. 
Therefore, it is derived from (\ref{e:0bG2ph}) by taking this limit:
\begin{eqnarray}
\frac{P_{\rm 0a}}{x_{0b}} \approx \lim_{x_{0b}\rightarrow 1}\{P_{0a}/x_{0b}\} = k_{\rm B}T \rho_{\rm b}
-\frac16 \rho_{\rm b}^2\int r\frac{dv_{00}(r)}{dr}g_{00}(r)d{\bf r} +\left.\frac{d\Delta}{dx_{1b}}\right|_{x_{1b}=0}\,,
\end{eqnarray}
which enables us to obtain Raoult's law due to the postulation $\left.{d\Delta}/{dx_{1b}}\right|_{x_{1b}=0}\!=\!0$, 
since the right-hand side of the above equation represents the vapor pressure of a pure liquid . On the other hand, Eq.~(\ref{e:1bG2ph}) leads to Henry's law as shown previously.

In some problem, the vapor pressure of a liquid can be neglected compared to the gas pressure $P_{\rm 1a}$ in the gas phase \cite{Murad,Pollock}, that is, the 0-species liquid is approximated to be non-volatile as is written in the form:
\begin{equation}\label{e:nonv1}
P_{\rm 0b} = k_{\rm B}T \rho_{\rm 0b}
-\frac16 \rho_{\rm 0b}^2\int r\frac{dv_{00}(r)}{dr}g_{00}(r)d{\bf r}
+\sum_{i\in {\rm 0b}}\langle{\bf r}_i\cdot{\bf F}_i^{\rm 0b} \rangle/3V_{\rm b}=\Gamma\,,
\end{equation}
since the non-volatility is defined by the relation: 
$P_0^{\rm w_{\rm a}}\!=\! P_{\rm 0b}-\Gamma\!=\!0$.
Therefore, the gas-liquid equilibrium is described by the following equation: 
\begin{equation}\label{e:eHenry}
P_{\rm 1a}=P_{\rm 1b}+\Gamma=k_{\rm B}T \rho_{\rm b}
-\frac16 \sum_{\rm \alpha,\beta}\rho_{\rm \alpha b}\rho_{\rm \beta b}\int r\frac{dv_{\alpha\beta}(r)}{dr}g_{\alpha\beta}(r)d{\bf r}\,,
\end{equation}
which is obtained by use of (\ref{e:Pvir-b}), (\ref{e:nonv1}) and (\ref{e:sumF01}).
Note that Eq.~(\ref{e:eHenry}) can be applied to determine the solubility $x_{\rm 1b}$ of a gas at any pressure $P_{\rm 1a}$ under the non-volatile approximation (\ref{e:nonv1}) in contrast with Henry's law with the constant (\ref{e:Henry}) which is only valid in the dilute solution.

It should be noted that the dilute-solution condition $\Delta\!=\!0$ appears both for  Raoult's law  and Henry's law as well as van't Hoff's law. This condition for the dilute limit (the zero solute-density limit) is satisfied 
because the term $\Delta$ is described in terms of interatomic forces ${\bf f}_{ij}$ between different kinds of species only as shown in (\ref{e:dilute}), that is, because the dilute limit is a pure-liquid limit where $\Delta$ becomes zero. 
Furthermore, we must postulate $\left.{d\Delta}/{dx_{1b}}\right|_{x_{1b}=0}\!=\!0$ 
or $\left.{d\Delta}/{dx_{0b}}\right|_{x_{0b}=0}\!=\!0$ for the two phase or osmotic system, respectively, to obtain the relations in the pure liquid limit; Symbols, $x_{1b}$ and $x_{0b}$, denote the solute concentration in the two phase or osmotic system, respectively. Therefore, the two different postulations, $\left.{d\Delta}/{dx_{1b}}\right|_{x_{1b}=0}\!=\!0$ 
and $\left.{d\Delta}/{dx_{0b}}\right|_{x_{0b}=0}\!=\!0$, express the same condition in the pure liquid limit.

In physiology, the partial pressure of water 
in solutions is important to see balance of water in body fluids. 
Water balance between \lq a' and \lq b' solutions separated by a membrane is maintained 
under the condition for the partial pressures of water: 
\begin{equation}\label{e:OSMwater}
P_{\rm water\!-\!a}=P_{\rm water\!-\!b}+\Gamma
\end{equation} 
with a proper pressure difference $\Gamma$; unless this condition is satisfied, the osmotic flow of water occurs. For example, when {\it M} kinds of impermeable solutes are dissolved in both a- and b-domains with partial pressures, $P_{i\rm a}$ and $P_{i\rm b}$, keeping the relation ${P}_i^{\rm w_{\rm b}}= P_{i\rm b}\!-\!P_{i\rm a}\!-\!\Gamma_i $, the above pressure difference is expressed by $\Gamma=\sum_{i=1}^{M}\Gamma_i$ as shown in Appendix~\ref{s:water}.

\section{Conclusion}\label{s:Con}
The purpose of this work is to prove the law of partial pressures in liquid mixtures, 
since at the present stage it is commonly believed that the law of partial pressures is valid only for ideal gases, that is to say, we can not define "partial pressures" for liquid mixtures. 
For this purpose, we note the law of partial pressures for osmotic systems, (\ref{e:mom}), (\ref{e:1Gam}) and (\ref{e:0Gam}), which are found by the computer experiment \cite{Itano} of osmotic systems, where "partial pressures" are observable by use of these relations.  Thus, we obtained the following results:

I) For the first time, we have defined "partial pressures" in liquid mixtures 
in terms of wall potentials confining each component to the same volume, 
that is, Eq.~(\ref{e:dPP1}) or (\ref{e:dPPbW}), which leads to Eq.~(\ref{e:Pvir1}) with (\ref{e:sumF01}). Using the definition of partial pressures (\ref{e:dPP1}), we have proven the law of partial pressures in liquid mixtures.

II) On the basis of partial pressures (\ref{e:dPP1}), from first principles we have proven the law of partial pressures for osmotic systems, (\ref{e:mom}), (\ref{e:1Gam}) and (\ref{e:0Gam}), which are found by the computer experiment. Also, this law is extended to osmotic systems consisting of many kinds of species as (\ref{e:OSMmks1}) and (\ref{e:OSMmks0}). Thus, water balance in body fluids, for example, can be described by (\ref{e:OSMwater}) in terms of the water "partial pressure".

III) Based on (\ref{e:dPP1}), (\ref{e:OSMmks1}) and (\ref{e:OSMmks0}), we have proven the fact that the gas-liquid phase equilibrium in mixtures  is established by each partial-pressure balance for every component in the gas- and liquid-phases, as is described in the relation: $P_{i\rm a}\!=\!P_{\i\rm b}\!+\!\Gamma_{i}$, that is (\ref{e:plinGL}). The non-volatility of $\alpha$-species in a solution is defined by $P_{\alpha\rm b}+\Gamma_\alpha$=0 on the basis of the definition of partial pressures.
From the gas-liquid phase equilibrium relation (\ref{e:plinGL}), Raoult's law and Henry's law (\ref{e:Henry}) with a formula to determine Henry's constant $\kappa_{\rm H}$ are derived for a dilute solution without use of chemical potentials, in conjunction with the solubility formula (\ref{e:eHenry}) of a gas at any gas pressure under the non-volatile approximation.
In addition, equation (\ref{e:plinGL}) has another face in physiology. There, a "dissolved gas tension" defined by $P_{i\rm a}\!=\!P_{i\rm b}+\Gamma_{i}$ plays an important role; a dissolved gas tension is essentially a partial pressure in liquid-solvent.

Our conclusion is summarized as follows. Until now, the law of partial pressures is supposed to be applicable only to ideal gases, and 
the definition of the partial pressures in liquid mixtures is taken to be insignificant because of some arbitrariness in the division of the total pressure into several parts. Nevertheless, we have shown in this investigation that the partial pressure can be defined uniquely as each wall pressure exerted by a component in the system [(\ref{e:dPP1})], and is an important observable physical quantity. As a consequence of this definition, the law of partial pressures is applicable to liquid mixtures with strong interactions as well as to ideal gases. 
Furthermore, it has been shown here that the partial pressures play an important role 
to see structures of the gas-liquid phase equilibrium in addition to the osmotic system: 
Henry's law,  Raoult's law and van't Hoff's law of osmotic pressure are confirmed as 
evidences that the partial pressures are important physical quantities. 
Also, the computer experiments \cite{Itano,Murad} support our definition of partial pressures (\ref{e:dPP1}).
\begin{acknowledgment}
We wish to thank Prof. H.~Kitamura for helpful and long-term discussions, which have clarified many problems in our work.
\end{acknowledgment}

\appendix

\section{Proof of (\ref{e:osm1})}\label{s:osmP}
In an osmotic system with the cubic volume as shown in Fig.\ref{fig-2}, three {\it external} forces, 
$F(0)$, $F(L/2)$ and $F(L)$, must be 
applied to the three surfaces, $S(0)$, $S(L/2)$ and $S(L)$, located at $x=0$, $x=L/2$ and $x=L$, respectively, to keep this osmotic system steady; 
\begin{equation}
 F(0){\bf e}_{\rm x} = [F(L/2)+F(L)]{\bf e}_{\rm x}\,, \label{e:PBL}
\end{equation}
with ${\bf e}_{\rm x}$ being the unit vector of the x-axis. 
Since $F(0)/L^2 = P_0^{\rm w}(0)+{P}_1^{\rm w}(0)=P_{\rm 0b}+P_{\rm 1b}$ and $F(L)/L^2 = P_1^{\rm w}(L)=P_{\rm 1a}$, we obtain $F(L/2)/L^2=P_{\rm 0b}+P_{\rm 1b}-P_{\rm 1a}\equiv P^{\rm osm}$.
%Note that this relation is valid even when the solvent molecules (1-species) interact with the semipermeable membrane $S(L/2)$, since  $F(L/2)/L^2$ is determined by  the interactions of molecules only with the walls, S(0) and S(L), irrespective of the membrane wall S(L/2).
On the other hand, we get $F(L/2)/L^2\equiv P^{\rm w}(L/2)= P_0^{\rm w_b}$ leading to (\ref{e:osm1}), since the wall potential caused by the membrane is assumed to interact only with the solute molecules in our model adopted for the osmotic system \cite{Itano,Crozier,RowleyI,RowleyII}.
In fact, in the standard treatment in MD simulation the osmotic pressure is determined by the wall pressure $P_0^{\rm w_b}$ rather than by the pressure difference for purposes of accuracy 
\cite{Itano,Crozier}, even when the solvent molecules interact with the semipermeable membrane \cite{Murad93,Murad,Rag}. The solvent-membrane interaction does not contribute to the wall pressure $P^{\rm w}(L/2)$ of the membrane to keep (\ref{e:osm1}) unchanged, since the solvent molecules move through the membrane in spite of the solvent-membrane interaction.

\section{Influence of wall potential with a finite range}\label{s:wall}

Thermodynamically, the wall potential $U^{\rm w}$ confining a fluid to a finite volume $V$ is assumed to be perfectly elastic and becomes abruptly infinite at the surface ${\partial V}$.
In reality the wall has short-range repulsive potentials which interact with inner particles to some range in the volume. Therefore, we obtain $P^{\rm w}\!=\!P\!=\!0$ if Eq.~(\ref{e:gVir}) is simply applied because of $P\!=\!0$ at the surface ${\partial V}$.  
The influence of the wall potential with a finite range has been examined by Green \cite{Green}, where he showed that at the surface $S'$ located at the distance $\Delta x$ from the wall, 
$P^{\rm w}\!=\!P$ is established and $P^{\rm w}\!=\!\int_0^{\Delta x}{dP(x)\over dx}dx$ 
for the one dimension case. 
Since the influence of the wall disappears at the surface $S'(=\!\partial V')$ located at the distance $\Delta x$ from the wall with an area $S(=\!\partial V)$, 
this influence is described as follows. Because the hydrostatic pressure $P(\bf r)$ is zero at the surface of the wall, there results,
$0=\oint_{S}\!{P}{\bf r}\cdot d{\bf S}=\Bigl[\oint_{S}-\oint_{S'}\Bigr]\!{P}{\bf r}\cdot d{\bf S}+\oint_{S'}\!{P}{\bf r}\cdot d{\bf S}=\int_{S\Delta x} \nabla \cdot\left({\bf r}{P}\right) d{\bf r}+\int_{V'} \nabla \cdot\left({\bf r}{P}\right) d{\bf r}$. 
Here, $-\int_{S\Delta x} \nabla \cdot\left({\bf r}{P}\right) d{\bf r}$ denotes the pressure effect exerted by the wall on the particles via a distortion in $P(\bf r)$ caused by the wall potential ${U}^{\rm w}$. Therefore, we obtain 
\begin{eqnarray}
\oint_{S}\!{P}^{\rm w}{\bf r}\cdot d{\bf S}
&=& -\int_{S\Delta x}  \nabla \cdot\left({\bf r}{P}\right) d{\bf r}=\int_{V'} \nabla \cdot\left({\bf r}{P}\right) d{\bf r}
= 3P(V\!-\!S\Delta x) \,, \label{e:Asx}
\end{eqnarray}
owing to $P(S')=P$. That is,
\begin{equation}
P^{\rm w}= P\left(1-\frac{ S\Delta x}{V}\right) \,.\label{e:therm1}
\end{equation}
The influence of the wall may disappear when the volume $V$ becomes infinite, and the pressure on the wall exerted by particles is taken as that given by the virial equation; this is the meaning of the virial equation (\ref{e:gVir2}) for a real wall potential.

When the pressure $P$ has a step-function like discontinuity on the surface $S(= \partial V_{\rm b})$ of the volume $V_{\rm b}$ involved in the volume $V$, a formula to calculate the surface integral is given by (\ref{e:avD}).
Here, we consider the case where the pressure $P$ increases continuously in the narrow domain between the two surfaces, ${S_{\!+\!\Delta x}}$ and ${S_{\!-\!\Delta x}}$ with $\Delta x$ being the order of the interatomic potential range, instead of a step-function like discontinuity on the surface $S$. Here, ${S_{\!+\!\Delta x}}$ and ${S_{\!-\!\Delta x}}$ denote a surface $S$ shifted outside by $\Delta x$ and that shifted inside, respectively.
In a similar way to obtain (\ref{e:therm1}), we can evaluate the surface integral $\oint_{\partial V}\!{P}{\bf r}\cdot d{\bf S}$ for this case on the basis of an identity:
$\oint_{\partial V}\!{P}{\bf r}\cdot d{\bf S}
=[\oint_{\partial V}\!{P}{\bf r}\!\cdot\! d{\bf S}-\oint_{S_{\!+\!\Delta x}}\!{P}{\bf r}\cdot d{\bf S}]+[\oint_{S_{\!+\!\Delta x}}\!{P}{\bf r}\cdot d{\bf S}-\oint_{S_{\!-\!\Delta x}}\!{P}{\bf r}\cdot d{\bf S}]
+\oint_{S_{\!-\!\Delta x}}\!{P}{\bf r}\cdot d{\bf S}$.
If we neglect quantities of the order ${ S\Delta x}/{V}$, we obtain from the above identity
\begin{eqnarray}
\oint_{\partial V}\!{P}{\bf r}\!\cdot\! d{\bf S}
&=& \oint_{\partial V_a}\!{P}{\bf r}\!\cdot\! d{\bf S}
 + \oint_{\partial V_b}\!{P}{\bf r}\!\cdot\! d{\bf S}
 + [P(S_{\!+\!\Delta x})\! -\! P(S_{\!-\!\Delta x}) ]\!\oint_{\partial V_b}\!\!\!{\bf r}\!\cdot\! d{\bf S} \,. \label{e:DisconTherm}
\end{eqnarray}

\section{Water balance in solutions with several solutes}\label{s:water}
A membrane separates solutions into \lq a' and \lq b' domains involving water (species-0) as a solvent. In solutions, {\it M} kinds of {\it impermeable} solutes are dissolved in both a- and b-domains with partial pressures, $P_{i\rm a}$ and $P_{i\rm b}$, keeping the relation ${P}_i^{\rm w_{\rm b}}= P_{i\rm b}\!-\!P_{i\rm a}\!-\!\Gamma_i $. In addition, ${\it M'}$ kinds of {\it permeable} solutes are dissolved in these domains with partial pressures, $\tilde P_{j\rm a}$ and $\tilde P_{j\rm b}$, keeping the relation ${P}_{j}^{\rm w_{\rm a}}=\tilde P_{j\rm a}=\tilde P_{j\rm b}\!+\!\tilde\Gamma_{j} $. Then, the osmotic pressure of this system is given by
$P^{\rm osm}\!=P^{\rm w_{\rm b}}\equiv\sum_{i=1}^M P_i^{\rm w_{\rm b}}=P_b-P_a$. Here, the total pressure of the domain-$\alpha$ ($\alpha\!=\!a$ or $b$) is defined by $P_\alpha \equiv P_\alpha^{\rm s}+P_{0\alpha}$ with  the solute pressure $P_\alpha^{\rm s}\equiv \sum_{i=1}^M\!P_{i\alpha}+\sum_{j=1'}^{M'}\!\tilde P_{j\alpha}$. Water balance is established under the condition 
${P}_{0}^{\rm w_{\rm a}}\!=\!P_{\rm 0a}\!=\!P_{\rm 0b}\!+\!\Gamma$ with $\Gamma\!=\!\sum_{i=1}^M \Gamma_{i}\!-\!\sum_{j=1'}^{M'}\tilde \Gamma_{j}$. The osmotic pressure is rewritten also as
${P}^{\rm w_{\rm b}}\!=\!P_{\rm b}^{\rm s}\!-\!P_{\rm a}^{\rm s}\!-\!\Gamma$. 

%\bibliography{LAWpp}

\begin{thebibliography}{10}



\bibitem{ChiharaEN}
J.~Chihara and M.~Yamagiwa: Prog. Theor. Phys. {\bfseries 118} (2007) 1019.

\bibitem{ChiharaUn}
J.~Chihara and M.~Yamagiwa: Prog. Theor. Phys. {\bfseries 111} (2004) 339.

\bibitem{Itano}
T.~Itano, T.~Akinaga, and M.~Sugihara-Seki: J. Phys. Soc. Jpn {\bfseries 77}
  (2008) 064605.

\bibitem{HOBBIE}
R.~K. HOBBIE: Proc. Nat. Acad. Sci. USA {\bfseries 71} (1974) 3182.

\bibitem{Woo}
K.~W. Woo and S.~I. Yeo: SNU J. Education Research {\bfseries 5} (1995) 127.

\bibitem{Schmidt-Nielsen}
K.~Schmidt-Nielsen: {\em Animal Physiology} (Cambridge Univ. Press, Cambridge,
  1997) 5th ed., p.~10.

\bibitem{Das}
D.~Das: {\em Biophysics and Biophysical Chemistry} (Academic Publishers, West
  Bengal, 2007), p.~50.

\bibitem{PP}
\url{http://www.princeton.edu/~achaney/tmve/wiki100k/docs/Partial_pressure.html}.

\bibitem{Stewart}
{Stewart's Textbook of Acid-Base}.
\url{http://www.acidbase.org/?show=sb&action=explode&id=67&sid=65}.

\bibitem{Willmer}
P.~Willmer, G.~Stone, and I.~Johnston: {\em Environmental Physiology of Animals} (
  Wiley, New York, 2009) second ed., p. 144.

\bibitem{Hoar}
W.~S. Hoar and D.~J. Randall: {\em Fish Physiology} (Academic Press, New York,
  1970), Vol.~4, p. 178.

\bibitem{Watten}
B.~J.~Watten, C.~E.~Boyd, M.~F.~Schwartz, S.~T.~Summerfelt, and B.~L.~Brazil: Aquacult. Eng {\bfseries 30} (2004) 83.

%\bibitem{Lion}
%T.~W.~Lion and R.~J.~Allen. : J. Chem. Phys. {\bfseries 137} (2013) 244911.

\bibitem{More79}
R.~M. More: Phys. Rev. A {\bfseries 19} (1979) 1234.

\bibitem{Chihara01}
J.~Chihara, I.~Fukumoto, M.~Yamagiwa, and H.~Totsuji: J. Phys.: Condens. Matter
  {\bfseries 13} (2001) 7183.

\bibitem{BaderAus}
R.~F.~W. Bader and M.~A. Austen: J. Chem. Phys. {\bfseries 107} (1997) 4271.

\bibitem{Bader}
R.~F.~W. Bader: J. Chem. Phys. {\bfseries 73} (1980) 2871.

\bibitem{RKubo}
R.~Kubo: {\em Statistical mechanics: an advanced course with problems and solutions} 
(North-Holland, Amsterdam 1971). p.~32.

\bibitem{Hansen}
J.~P. Hansen and I.~R. McDonald: {\em Theory of Simple Liquids} (Academic
  Press, London, 1987) second ed., p.~20.

\bibitem{Murad93}
S.~Murad and J.~G. Powles: J. Chem. Phys. {\bfseries 99} (1993) 7271.

\bibitem{Murad}
S.~Murad and S.~Gupta: Chem. Phys. Lett. {\bfseries 319} (2000) 60.

\bibitem{Pollock}
T.~R. Pollock, P.~Crozier, and R.~L. Rowley: Fluid Phase Equilb. {\bfseries
  217} (2004) 89.

\bibitem{Crozier}
P.~S. Crozier: Dissertation, Brigham Young University (2001).\\
 p.49. http://contentdm.lib.byu.edu/cdm/ref/collection/ETD/id/1.

\bibitem{RowleyI}
R.~L. Rowley, T.~D. Shupe, and M.~W. Schuck: Mol. Phys. {\bfseries 82} (1994)
  841.

\bibitem{RowleyII}
R.~L. Rowley, M.~W. Schuck, and J.~C. Perry: Mol. Phys. {\bfseries 86} (1995)
  125.

\bibitem{Rag}
A.~V. Raghunathan and N.~R. Aluru: Phys. Rev. Lett. {\bfseries 97} (2006)
  024501.

\bibitem{Green}
H.~S. Green: {\em The Molecular Theory of Fluids} (North-Holland, Amsterdam,
  1952), p.~206.

\end{thebibliography}
%\bibliographystyle{jpsj}

%\begin{verbatim}
%\profile{Taro Butsuri}{was born in Tokyo, Japan in 1965. ...}
%\end{verbatim}

%%\begin{thebibliography}{9}
%%\bibitem{jpsj} The abbreviation for JPSJ must be ``J. Phys. Soc. Jpn." \note{in the reference list}.
%%\bibitem{instructions} More abbreviations of journal titles are listed in ``Instructions for Preparation of Manuscript".
%%\end{thebibliography}

\end{document}